%% file: main.tex
\documentclass{article}
\usepackage[english]{babel}

\usepackage[letterpaper,top=2cm,bottom=2cm,left=3cm,right=3cm,marginparwidth=2.cm]{geometry}

\usepackage[sorting=none]{biblatex}
\usepackage{amsmath}
\usepackage{graphicx}
\usepackage[colorlinks=true, allcolors=blue]{hyperref}
\usepackage{multirow}
\usepackage{authblk}
\usepackage{doi}
\usepackage[colorinlistoftodos]{todonotes}
\usepackage{xspace}
\usepackage{mydefinitions}
\usepackage{caption}
\usepackage{subcaption}

\title{The k4Clue package: \\ Empowering Future Collider Experiments \\ with the CLUE Algorithm}
\author[1]{E Brondolin\footnote{Corresponding author.\\
E-mail addresses: erica.brondolin@cern.ch (E. Brondolin)}}
\author[1]{M Rovere}
\author[1]{F Pantaleo}
\affil[1]{CERN, Switzerland}

\begin{document}
\maketitle

\begin{abstract}
High granularity calorimeters have become increasingly crucial in modern particle physics experiments, and their importance is set to grow even further in the future. The CLUstering of Energy (CLUE) algorithm has shown excellent performance in clustering calorimeter hits in the High Granularity Calorimeter (HGCAL) developed for the Phase-2 upgrade of the CMS experiment. In this paper, we investigate the suitability of the CLUE algorithm for future collider experiments and test its capabilities outside the HGCAL software reconstruction. To this end, we developed a new package, k4Clue, which is now fully integrated into the Gaudi software framework and supports the EDM4hep data format for inputs and outputs. We demonstrate the performance of CLUE in three detectors for future colliders: CLICdet for the CLIC accelerator, CLD for the FCC-ee collider and a second calorimeter based on Noble Liquid technology also proposed for FCC-ee. We find excellent reconstruction performance for single gamma events, even in the presence of noise, and also compared with other baseline algorithms. Moreover, CLUE demonstrates impressive timing capabilities, outperforming the other algorithms and independently of the number of input hits. This work highlights the adaptability and versatility of the CLUE algorithm for a wide range of experiments and detectors and the algorithm's potential for future high-energy physics experiments beyond CMS.
\end{abstract}

\tableofcontents

\section{Introduction}\label{intro}
The High Energy Physics field is transitioning towards using calorimeters with higher lateral and longitudinal readout granularity due to their numerous advantages, such as providing more precise particle identification, detailed reconstruction of electromagnetic and hadronic showers, and improved separation of signals from pile-up events. 

In the near future, the CMS High Granularity Calorimeter (HGCAL), developed for the HL-LHC era, will represent the next step in the generation of high-granularity calorimeters~\cite{TDR_HGCAL}. HGCAL is a calorimeter that boasts exceptional fine transverse and longitudinal segmentation for both electromagnetic and hadronic compartments. The HGCAL design utilizes silicon sensors as active material in the front sections, with readout cells of approximately $1\cm^{2}$ area, while the back of the detector features a combination of silicon sensors and scintillator tiles. Other calorimeters with similar technology and granularity are also being explored for prototype detectors at future linear and circular electron-positron colliders.

To address the computational challenge posed by the large data scale, the CLUE (CLUstering of Energy) algorithm was developed in the context of HGCAL. Its CPU and GPU standalone implementations have demonstrated its performance, highlighting the significance of algorithmic parallelization in the coming era of heterogeneous computing~\cite{CLUE}. 

This paper is organized as follows: Section~\ref{sec:clue} provides an overview of the latest developments in CLUE and its input parameters, while Section~\ref{sec:k4clue} outlines its implementation within the future collider software framework and describes several improvements made to the baseline version. Finally, Section~\ref{sec:results} demonstrates CLUE's capabilities for several electromagnetic calorimeters at future collider experiments.
\input{clue.tex}
\input{k4clue.tex}
\input{results.tex}


\section{Conclusions}\label{conclusions}
In conclusion, the development of the \texttt{k4CLUE} package has enabled the use of the CLUE algorithm within the \texttt{Key4hep} software stack. The package has improved upon the standalone CLUE package, allowing for use on the full detector and different types of calorimeters. Analysis on three different future calorimeters has demonstrated the good performance of the CLUE algorithm for single gamma events, even in the presence of noise, and compared favorably to other baseline algorithms. This work highlights not only the adaptability and versatility of the CLUE algorithm for a wide range of experiments and detectors but also the excellent timing capabilities.

\section*{Acknowledgments}
The authors express their gratitude to CERN, Switzerland, for ongoing support. This research was supported by the CERN Strategic R\&D Programme on Technologies for Future Experiments~\cite{EP_RnD}. Special thanks go to Brieuc Francois for providing support on the Noble Liquid Calorimeter results, as well as the entire key4hep team for their valuable assistance throughout the development of the \texttt{k4Clue} package. In particular, the authors would like to thank Andre Sailer, Frank Gaede, Thomas Madlener, Placido Fernandez Declara, and Valentin Volks for their contributions and discussions.

\begin{figure}[h]
\begin{center}
\includegraphics[width=0.70\textwidth]{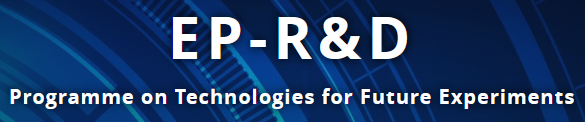}
\end{center}
\end{figure}


\printbibliography

\end{document}

%% file: clue.tex
\section{CLUE}\label{sec:clue}

CLUE (CLUstering of Energy)\cite{CLUE} is a fast and innovative clustering algorithm designed for grouping digitized energy deposits (hits/points) left by particles traversing the active sensors in 2D clusters. It is inspired by the Clustering by Fast Search and Find Density Peak algorithm~\cite{CFSFDP} and is primarily used for high-granularity sampling calorimeters. The algorithm operates layer-by-layer to group the hits and creates 2D clusters with a well-defined \textit{seed} hit and a list of \textit{follower} hits assigned to it. It also identifies \textit{outliers} that do not belong to any clusters. 

The algorithm is fully parallelizable and can be optimized using hardware accelerators such as graphics processing units (GPU). The algorithm's code is available as a standalone repository in~\cite{CLUE_gitlab}, and it has undergone updates and parameter definition revisions since its initial publication.

\subsection{Clustering procedure}\label{sec:procedure}

To enable efficient neighbor search in CLUE, a fixed-grid spatial index is constructed as a first step. This index organizes the hits into bins based on their 2D coordinates. When searching for neighbors within a specified distance $d$, CLUE only examines hits within the bins touched by a window of size $(x_i \pm d, y_i \pm d)$ centered on each point of interest ($N_{d}$). This approach is advantageous because the space division is independent of any particular distribution of data points and thus can be explicitly predefined and easily parallelized.

In the second step of the algorithm, the hit's local energy density $\rho$ and its distance $\delta$ to the nearest hit with higher local density are calculated using the hits position and energy value.  

To compute $\rho$, all neighbors within a specified distance $d_c$ are taken into account. The equation for $\rho_i$ is given as
\begin{equation}
    \rho_i = \sum_{j:j \in N_{d_c}(i)} \chi (d_{ij}) w_{j}
\end{equation}
where $w_j$ is the energy of point $j$ and the convolution kernel $\chi_{ij}$ is defined such that it is equal to 1.0 when $i=j$, 0.5 when $0 \leq d_{ij} \leq d_c$, and 0.0 when $d_{ij} > d_c$. 
After computing $\rho$, the nearest-higher hit is defined as the closest hit with the higher local energy density among the points in a space with dimension $N_{dm}$, where $d_m = o_f \times d_c$ and $o_f$ is the outlier delta factor.

In the last step, each hit is promoted as a seed or demoted as an outlier according to the following criteria:
\begin{itemize}
    \item  seed: $\rho \geq \rho_c$ and $\delta \geq d_c$;
    \item outlier: $\rho < \rho_c$ and distance $\delta \geq ( o_f \times d_c )$.
\end{itemize}
where $\rho_c$ is the minimum energy density to promote a point as a seed or the maximum density to demote a point as an outlier, 
$d_c$ is also called critical distance, and $o_f$ the outlier delta factor parameter to define
the maximum distance for a hit to be linked to the nearest higher point.
After having defined seeds and outliers, CLUE registers each remaining point as a follower to its nearest-higher and then it collects all the hits belonging to a cluster by passing cluster indices from the seeds to their followers, iteratively. 
The expansion of each cluster is independent of all the others. 

The choice of the values of the three input parameters ($\rho_c$, $d_c$, and $o_f$) should be based 
on the detector's characteristics and physics objects. For example, the
$\rho_c$ parameter can be adjusted to lower the likelihood of forming clusters solely due to noise contributions;
and $d_c$ and $o_f$ can be chosen based on the shower sizes, their separations, and the granularity of the detector.

CLUE has demonstrated its clustering capabilities in various scenarios, including synthetic datasets~\cite{CLUE} and simulated deposited energy in the CMS HGCAL detector's individual cells. In a study presented in~\cite{clue_acat2021}, CLUE's performance was found to be excellent in terms of energy response and its ability to withstand noise for different particle types interacting in the HGCAL calorimeter.

%% file: k4clue.tex
\section{The \texttt{k4Clue} package}\label{sec:k4clue}

\subsection{Integrating CLUE in \texttt{Key4hep}}\label{sec:key4hep}
The optimization process of future collider experiments requires maximal flexibility in detector geometries, 
materials and sensitive areas, and efficient simulation and analysis tools to quantify the overall performance. 
For this purpose, the commissioning of a common ‘Turnkey Software Stack’ (\texttt{Key4hep}) has been a high-priority project
among all major future collider projects~\cite{Key4hep_article2021, key4hep_chep2021}. 
It aims to design a common set of software tools for detector optimization 
and physics performance studies to cover most, if not all, future linear and 
circular machines colliding leptons and hadrons.

An integrated solution for detector simulation based on the \texttt{iLCSoft}~\cite{ilcsoft_github} and \texttt{FCCSW}~\cite{fccsoft_github} frameworks 
is currently being developed using the following three main ingredients:
\begin{itemize}
    \item a new common event data model, \texttt{EDM4hep}~\cite{edm4hep_github}, combining together the features of the \texttt{PODIO} EDM-toolkit and the most common event data models used by the linear and the circular collider communities, \texttt{LCIO} and \texttt{FCC-EDM}, respectively;
    \item an event processing framework, \texttt{Gaudi}~\cite{gaudi_v33r0}, that has already a large user and developer community in the LHC experiments and offers - or plans to offer - support for access to heterogeneous resources and task-oriented concurrency; 
    \item a package manager, \texttt{Spack}~\cite{spack}, which provides a recursive specification syntax to invoke builds of packages and their dependencies, regardless of the environment and the different sources, to enable the installations of the \texttt{Key4hep} software on a variety of platforms.
\end{itemize}
The documentation to start using the \texttt{Key4hep} software is available in~\cite{key4hep_website}.

The development of a new package, called \texttt{k4CLUE}, was necessary to enable the use of the CLUE algorithm within the \texttt{Key4hep} stack~\cite{k4clue_v01-00-03}.  
This package, specifically \texttt{k4CLUE (v01-00-03)}, provides a wrapper class that allows for the execution of CLUE in the \texttt{Gaudi} software framework, with \texttt{EDM4hep} data format being used for both inputs and outputs. In addition, some modifications were made to the original algorithm and its underlying data structures to enhance its clustering capabilities and to enable testing on a variety of future experiments.

Currently and in parallel to this work, the CLUE standalone repository is being refactored to incorporate the most recent updates introduced in \texttt{k4CLUE} and further enhancements. The ultimate goal is to adapt it to work as an external library and integrate it directly into the \texttt{Key4hep} software stack in the near future.

\subsection{\texttt{k4CLUE} additional features}\label{sec:k4clue_additional}

As mentioned in Sec.~\ref{sec:procedure}, CLUE uses a fixed-grid spatial index to access and query spatial data points efficiently. 
Thus, a multi-layer tessellation is created which divides the 2D space into fixed rectangular bins (\texttt{LayerTile}). 
The limits of the tessellated space and tiles dimensions are defined by the user.

The first area for developments we identified in the CLUE standalone package was the definition of the tessellated space.
The \texttt{LayerTile} in the standalone version defines coordinates and searches only in the transverse plane, 
and thus it assumes that the \texttt{LayerTile} of the calorimeter is placed in the forward region. 
To cluster hits in the entire detector region, the basic structure of the \texttt{LayerTile} and the search algorithm had to be modified
to allow for the definition of a cylindrical surface, typical of particle detectors placed in the barrel region.

In the case of a cylindrical \texttt{LayerTile}, the $(x,y)$ coordinates defined in CLUE change meaning:
\begin{equation}
    x \rightarrow r\phi \qquad \qquad y \rightarrow z
\end{equation}
where $r$ is the radius of the cylindrical layer.

The second area for improvement that we identified was the need to modify the data structures 
employed in the algorithm to accommodate various calorimeter layouts.
A dedicated documentation page in the package (\texttt{include/readme.md})
allows the user to follow a simple but detailed step-by-step procedure to introduce and test the preferred layout.

\subsection{Testing and validation}\label{sec:k4clue_validation}

\texttt{k4Clue} uses the GitHub continuous integration process to ensure that the modifications
or additions to the software do not break the clusterization process. 
In the latest release tests are performed on C++ code and using \texttt{EDM4hep} data.

The \texttt{k4Clue} package comes with a dedicated class called \texttt{clue::CLUECalorimeterHit}, which extends the existing \texttt{EDM4HEP::CalorimeterHit} class. This class has additional methods that are specific to the CLUE algorithm, such as identifying the detector region and classifying the hit as a seed, follower, or outlier. 
Utilizing this class allows for direct performance studies on the \texttt{EDM4HEP} collection, offering a notable advantage.
Additionally, a dedicated analysis code is provided with \texttt{k4CLUE} that produces \texttt{ROOT::Ntuples}, 
allowing for further testing and visualization of the calorimeter hits belonging to the cluster.

%% file: results.tex
\section{Performance evaluation}\label{sec:results}

We showcase the clustering performance of \texttt{k4CLUE} using various types of highly granular electromagnetic calorimeters developed for future $e^+e^-$ colliders that aim to fulfill the demands of the particle flow paradigm and offer excellent photon energy resolution over a broad energy spectrum. These detectors are highly granular and some of them use similar technology to the HGCAL detector, making them ideal candidates to evaluate the capabilities of \texttt{k4CLUE}.

As a first step, we assess the performance of \texttt{k4CLUE} for the highly granular electromagnetic calorimeters proposed 
in the CLICdet detector for the CLIC accelerator~\cite{CLICdet} 
and the CLD detector (CLIC-like detector) for the FCC-ee accelerator~\cite{CLD}.
Both detectors are sampling calorimeters of 40 layers of $5 \times 5 \mm^{2}$ silicon sensors interspersed by tungsten plates in the barrel and in the endcap region. 
The primary difference between the CLICdet and CLD calorimeters is in their detector layout parameters. Specifically, due to the larger tracking area required for the CLD experiment to compensate for a lower detector solenoid field, the CLD design starts from a larger radius in both the barrel and endcap regions compared to the CLICdet design.

A third type of calorimeter, based on noble liquid technology, was also included in the \texttt{k4CLUE} testing. This technology has been proposed as the baseline scenario for an FCC-hh experiment due to its radiation hardness and excellent energy resolution and is currently being investigated for the barrel region of one of the detector designs of the $e^+e^-$ collider as well~\cite{LAr_FCChh_2019, LAr_FCCee_2022}. The sampling detector selected for testing \texttt{k4CLUE} features inclined Lead planes used as absorber material interleaved with gaps filled with liquid Argon (LAr), where the signal is induced. The detector granularity is determined by the readout cells of the electrodes situated in the middle of the noble liquid gap. The cell size in phi increases radially, ranging between $17.9$ and $20.7\mm$, while in eta it is about $20\mm$. 
A total of $12$ longitudinal compartments are simulated.

A summary of the main key parameters of the three electromagnetic calorimeters is given in Table~\ref{tab:calo}.

\begin{table}[]
    \centering
    \begin{tabular}{l|c|c|c}
         & CLD & CLICdet & LAr\\
        \hline  
        ECAL technology & Silicon/W & Silicon/W & LAr/Pb\\
        ECAL radiation lengths [$X_0$] & 22 & 22 & 22 \\
        \hline
        ECAL barrel $r_{min}$ [mm] & 1500 & 2150 & 2100\\
        ECAL barrel $r_{max}$ [mm] & 1702 & 2352 & 2770\\
        ECAL barrel $z_{max}$ [mm] & 2210 & 2210 & 3100\\
        \hline
        ECAL endcap $r_{min}$ [mm] & 410 & 340 & -\\
        ECAL endcap $r_{max}$ [mm] & 1700 & 2455 & -\\
        ECAL endcap $z_{min}$ [mm] & 2307 & 2307 & -\\
        ECAL endcap $z_{max}$ [mm] & 2509 & 2509 & -       
    \end{tabular}
    \caption{Comparison of key parameters of the different electromagnetic calorimeters for future colliders.}
    \label{tab:calo}
\end{table}

\input{cld.tex}

\input{clicdet.tex}

\input{noble_liquid_calo.tex}

\input{computing_time.tex}

%% file: cld.tex
\subsection{Performance for CLD and CLICdet detector}\label{sec:cld_clicdet}

The detector simulation and reconstruction studies presented in this work for CLD and CLICdet were conducted using the \texttt{CLICPerformance} package, which was developed within the \texttt{ILCSoft} and modified to operate in the \texttt{Key4hep} Framework through the use of the \texttt{k4MarlinWrapper} package~\cite{k4MarlinWrapper_v00-04-02}.

To ensure that neighboring cells are contained, the granularity of the detector's geometry is taken into account to determine the critical distance of CLUE, $d_c$. Therefore, both in the case of CLD and CLICdet, $d_c$ is set to three times the size of a single silicon cell, resulting in $d_c = 15.0\mm$. The tuning parameters for the outlier delta factor, $o_f$, and the minimum local density, $\rho_c$, were tested with different values using 500 events of a single gamma generated on the inner surface of the calorimeter with a monochromatic energy of 10\GeV in the case of the CLD calorimeter. Similar conclusions can be applied to the case of the CLICdet calorimeter.

\subsubsection{CLD calorimeter results}\label{sec:cld}

\begin{figure}
     \centering
     \begin{subfigure}[b]{0.45\textwidth}
         \centering
         \includegraphics[width=\textwidth]{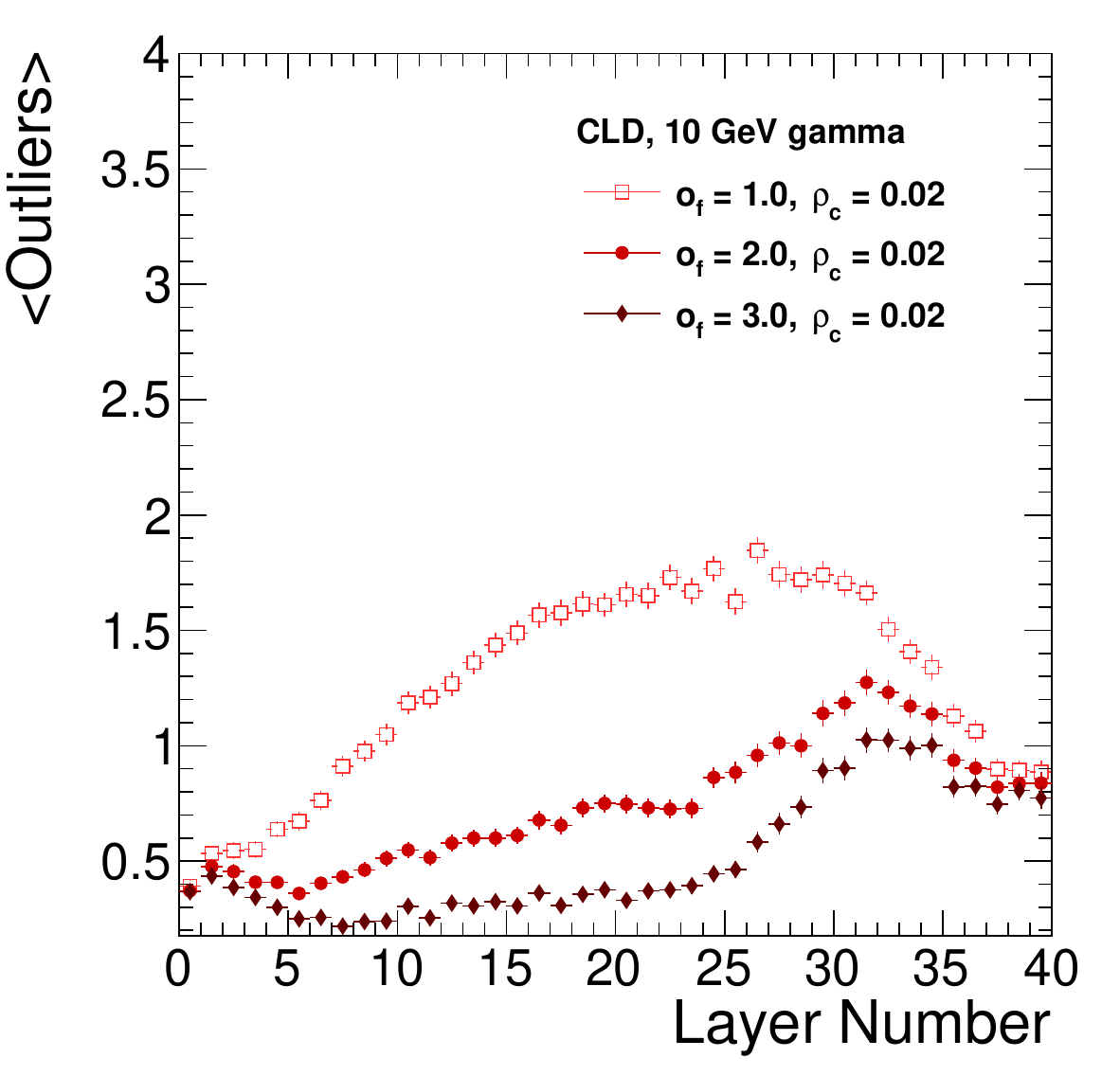}
     \end{subfigure}
     \hfill
     \begin{subfigure}[b]{0.45\textwidth}
         \centering
         \includegraphics[width=\textwidth]{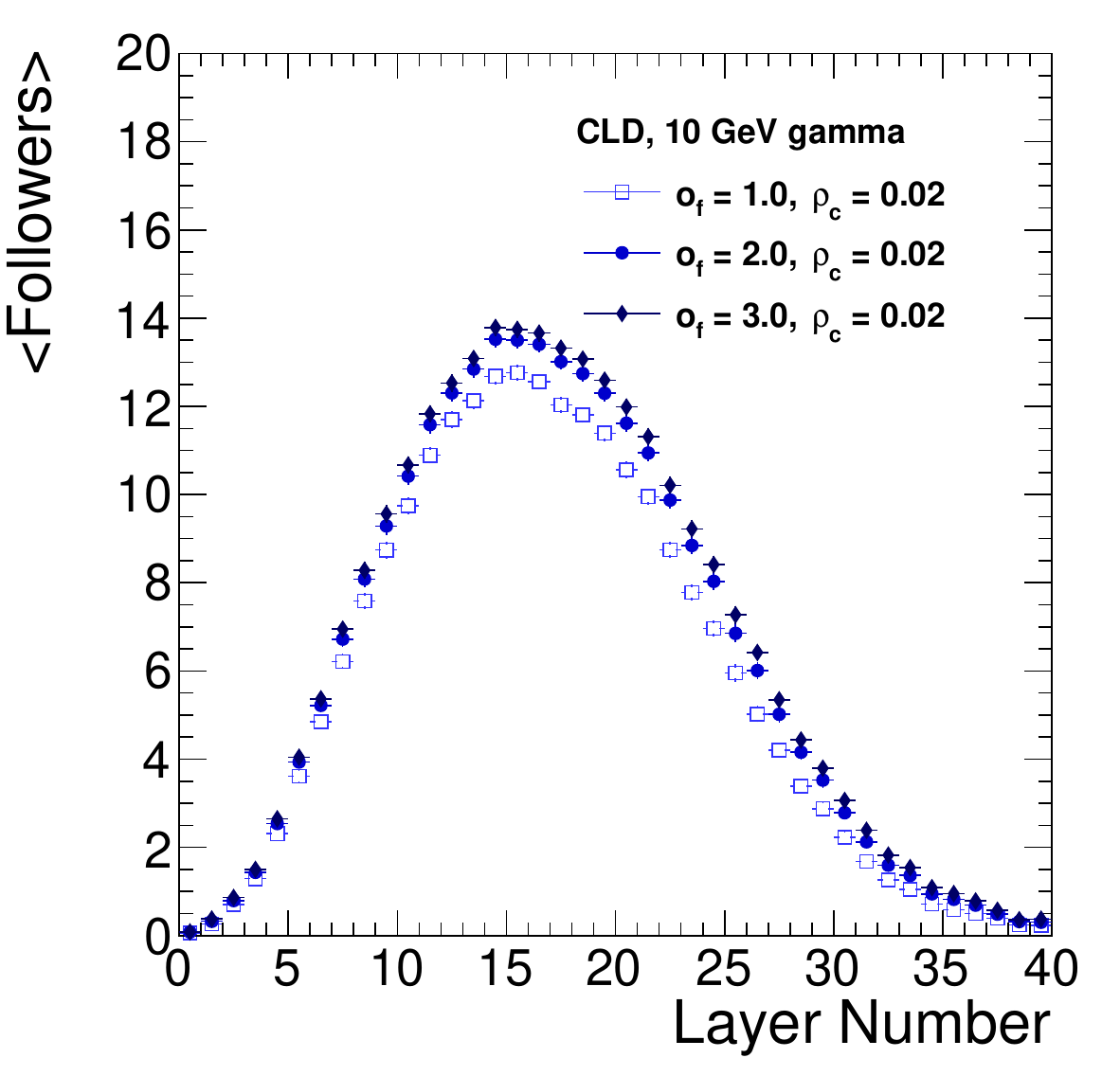}
     \end{subfigure}
     \newline
     \hfill
     \begin{subfigure}[b]{0.45\textwidth}
         \centering
         \includegraphics[width=\textwidth]{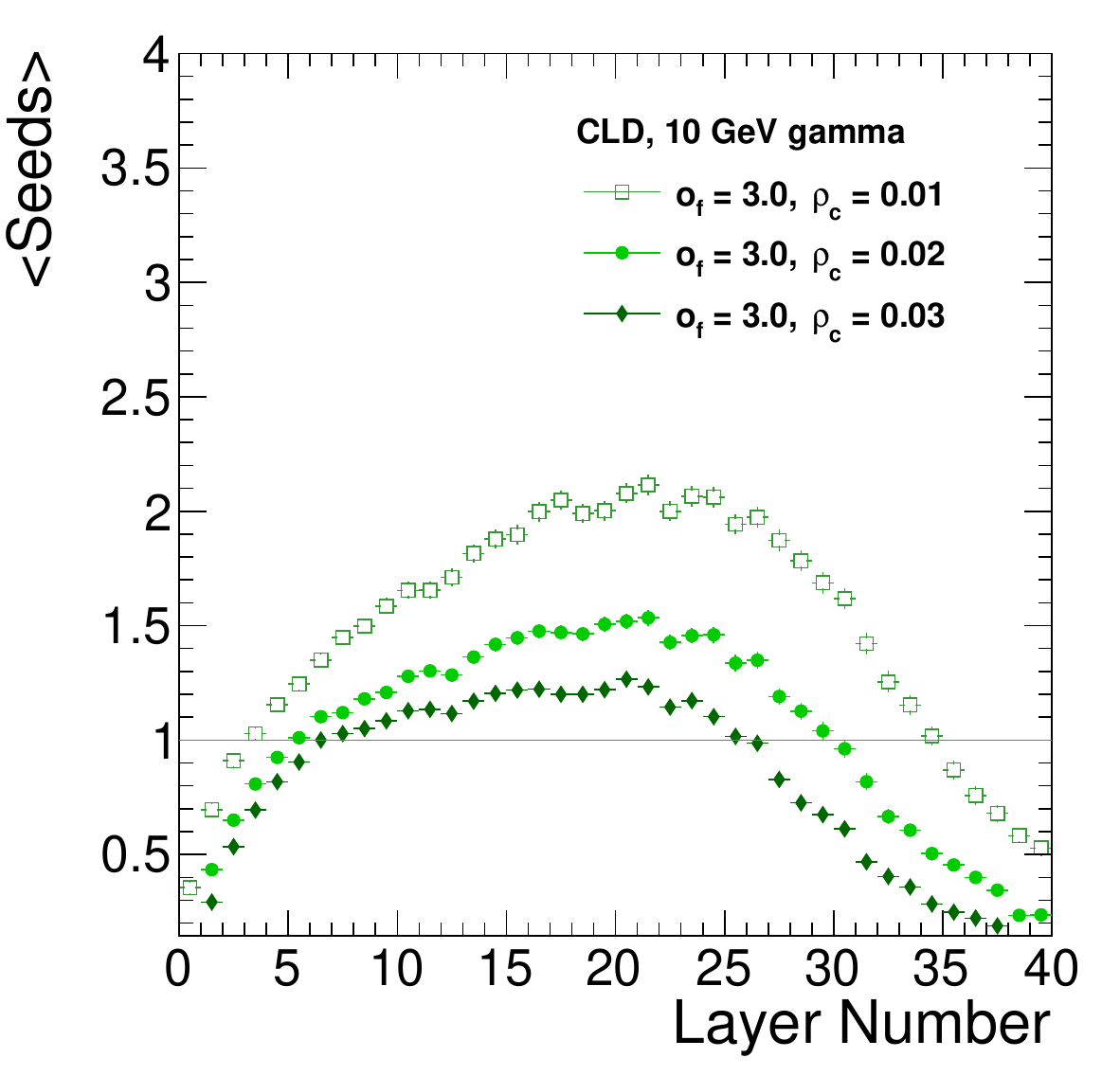}
     \end{subfigure}
     \hfill
     \begin{subfigure}[b]{0.45\textwidth}
         \centering
         \includegraphics[width=\textwidth]{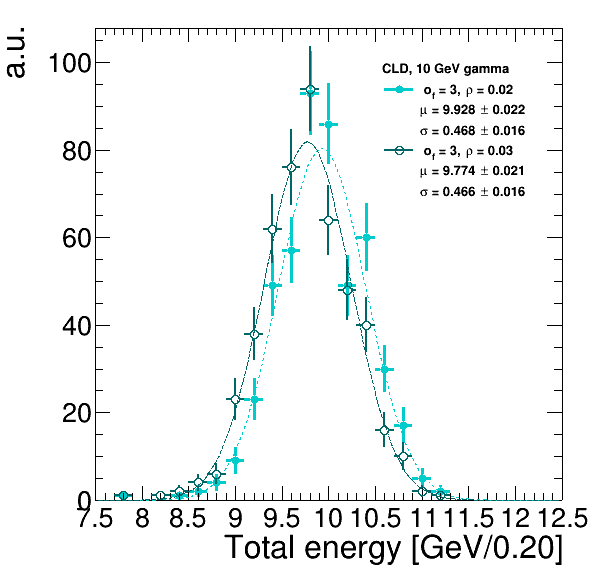}
     \end{subfigure}
     \caption{Top: Average number of outliers (left) and followers (right) as a function of the layer number
for three different values of the outlier delta factor ($o_f = 1, 2, 3$). The $\rho_c$ parameter is kept constant at $0.02$.\\
Bottom: In the left plot, the average number of seeds is shown as a function of the layer number for three different values of the critical density ($\rho_c = 0.01, 0.02, 0.03$). In the right plot, the distribution of the energy clustered by CLUE is shown for $\rho_c = 0.02, 0.03$. The $o_f$ parameter is kept constant at $3$. \\
Results are produced with single gamma events generated from the inner surface of the CLD calorimeter with monochromatic energy of 10\GeV.
}
     \label{fig:cld_tuningPars}
\end{figure}

 The results for the tuning parameters study are shown in Figure~\ref{fig:cld_tuningPars}. The top row of the figure displays the average number of outliers (left) and followers (right) per layer for three outlier delta factor values ($o_f = 1, 2, 3$). The layer number is plotted along the horizontal axis. The follower profile is minimally influenced by this parameter choice, whereas the outlier distribution shows a clear dependence. For the largest value of $o_f$, the number of outliers in the shower is considerably diminished and is primarily concentrated in the final few layers of the calorimeter. This is due to the fact that the hits corresponding to the electromagnetic shower are less energetic in these layers.

This finding is further supported by the seed distribution presented in the lower left panel of the figure for different values of the critical density ($\rho_c = 0.01, 0.02, 0.03$). While using $\rho_c = 0.03$ brings the number of seeds closer to one per layer, the final input parameter choice should consider the total energy distribution collected in the clusters as well. This distribution is shown in the lower right panel of the figure for $\rho_c = 0.02$ and $0.03$ with $o_f$ fixed at $3$, and a Gaussian function is fit to the distribution. The width and sigma parameters of the Gaussian are reported in the plot, and the obtained sigma value ($0.047\%$ at $10\GeV$) is consistent with the energy resolution results for single gamma presented in~\cite{CLD}.

\begin{figure}
     \centering
     \begin{subfigure}[c]{0.45\textwidth}
         \centering
         \includegraphics[width=\textwidth]{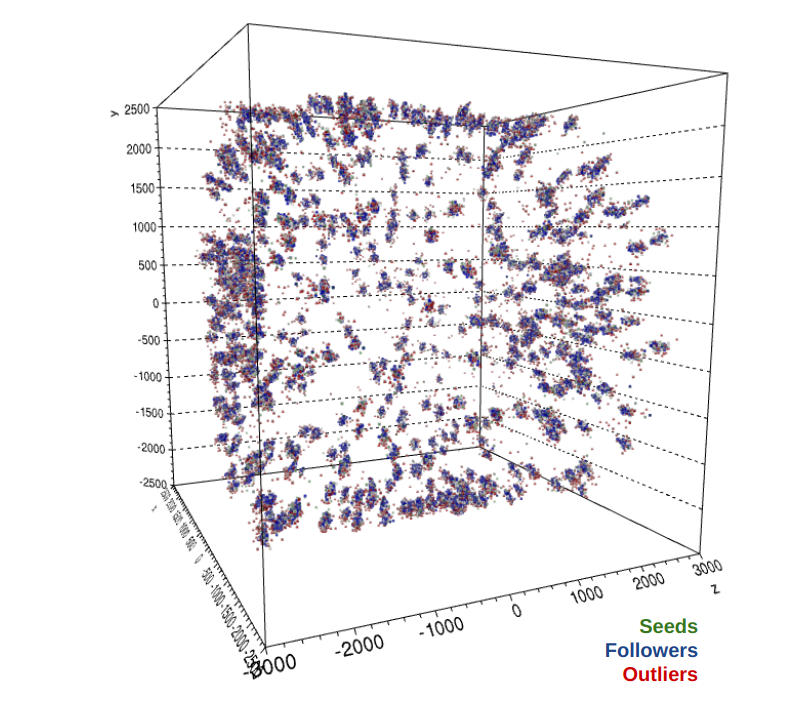}
         \label{fig:cld_multipleGammas_clueDisplay}
     \end{subfigure}
     \begin{subfigure}[c]{0.40\textwidth}
         \centering
         \includegraphics[width=\textwidth]{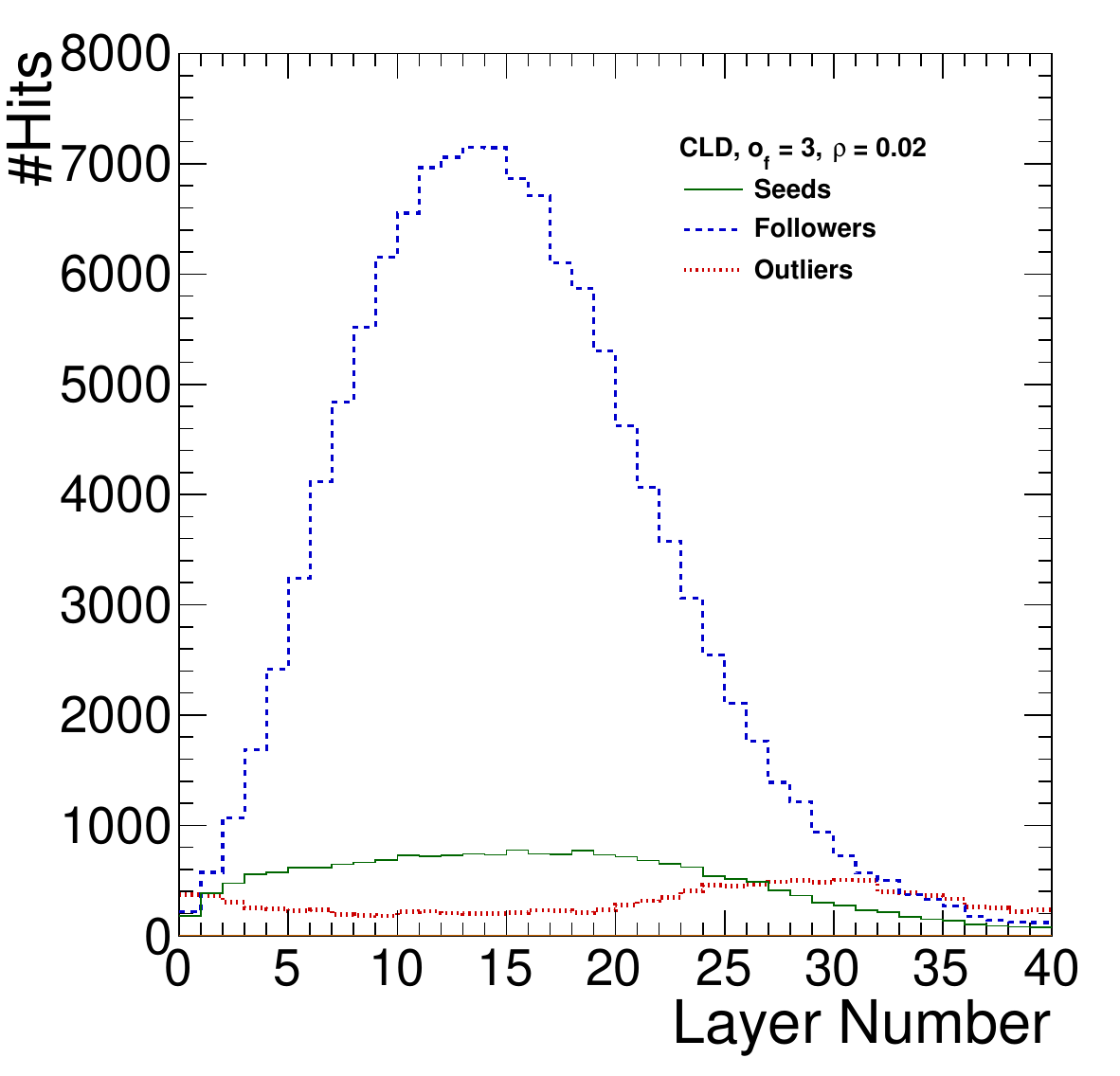}
         \label{fig:cld_multipleGammas_profile}
     \end{subfigure}
     \caption{
     Total number of followers (dashed line), seeds (solid line) and outliers (dotted  line) as a function of the layer number
     for a single event comprising 500 gammas generated at the same time from the inner surface of the CLD calorimeter 
     with monochromatic energy of 10\GeV. The CLUE input parameters used are $o_f = 3$ and $\rho_c = 0.02$.}
     \label{fig:cld_multipleGammas}
\end{figure}

CLUE's performance in more complex environments was evaluated by simulating a single event containing 500 gamma particles, 
with input parameters set at $o_f = 3$ and $\rho_c = 0.02$ based on the considerations discussed earlier.
In Figure~\ref{fig:cld_multipleGammas}, the hit distribution as a function of the layer number is shown 
for CLUE's followers, seeds, and outliers. 
The distribution closely resembles the ones observed for the single gamma case, 
and the number of seeds in the core of the shower (Layer 10-25) is between 1 or 2 times the number of events, 
indicating similar behavior as above. 
These results confirm CLUE's capability to reconstruct showers even in more challenging event scenarios.

%% file: clicdet.tex
\subsubsection{CLICdet calorimeter results}\label{sec:clicdet}

Since the CLIC calorimeter has similar characteristics to the CLD calorimeter, the same input values for CLUE ($\rho_c = 0.02$ and $o_f = 3$) are used to evaluate its performance at different energies. In this regard, a subset of the results obtained for single gamma events at energies of 10\GeV and 100\GeV is presented here.

\begin{figure}
     \centering
     \begin{subfigure}[b]{0.45\textwidth}
         \centering
         \includegraphics[width=\textwidth]{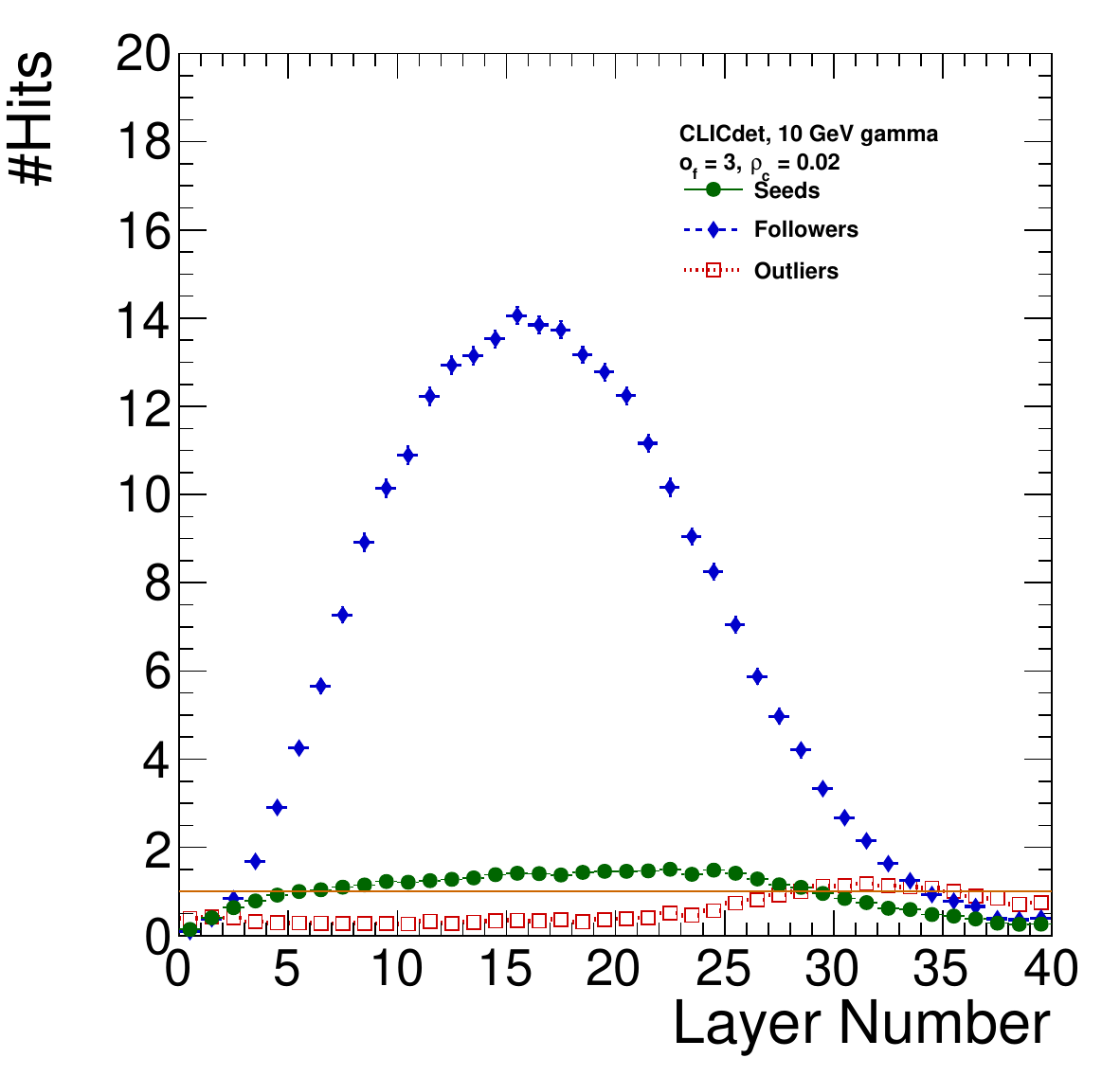}
         \label{fig:clicdet_profile_10GeV}
     \end{subfigure}
     \hfill
     \begin{subfigure}[b]{0.45\textwidth}
         \centering
         \includegraphics[width=\textwidth]{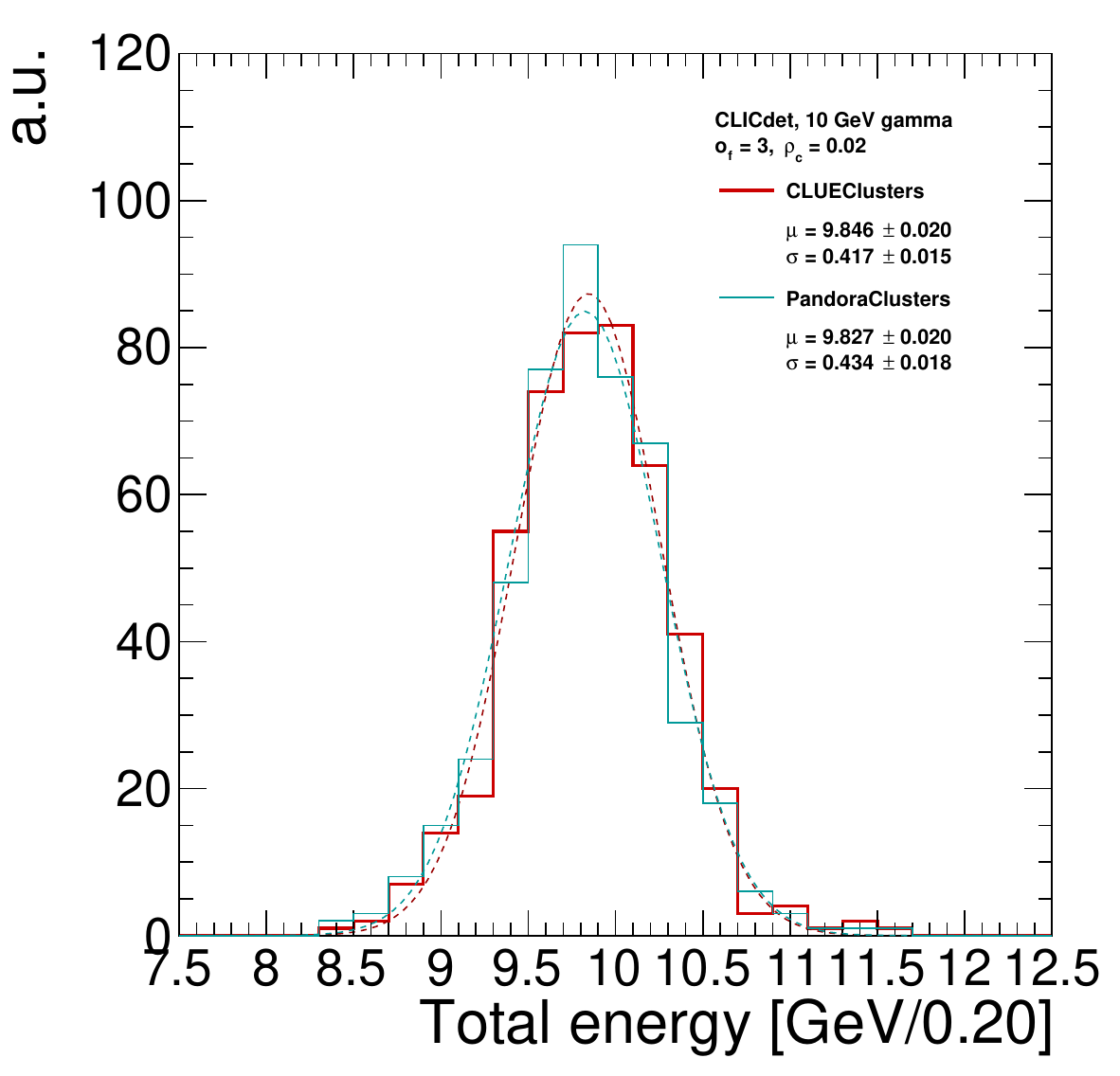}
         \label{fig:clicdet_energy_10GeV}
     \end{subfigure}
     \\
     \begin{subfigure}[b]{0.45\textwidth}
         \centering
         \includegraphics[width=\textwidth]{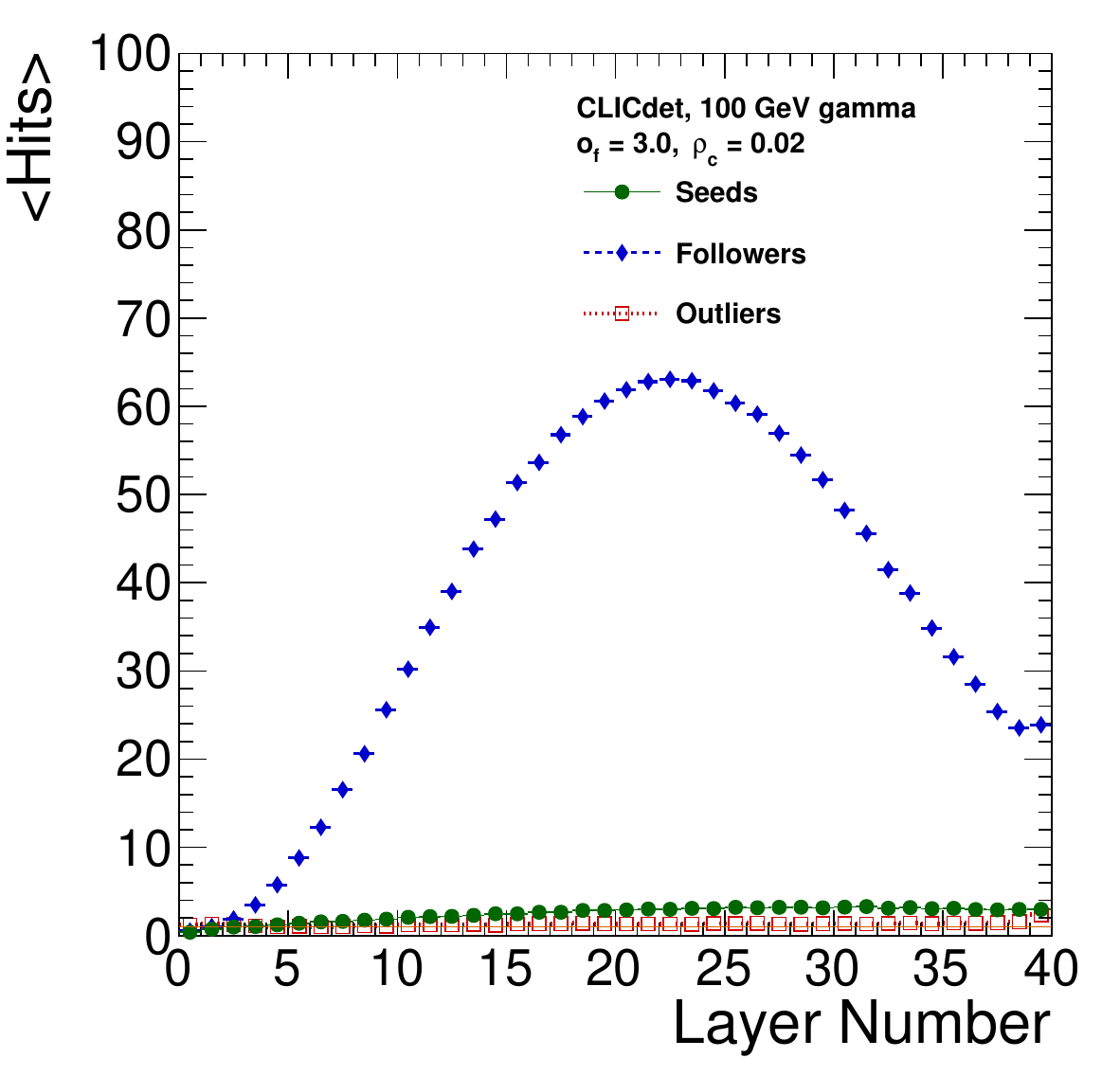}
         \label{fig:clicdet_profile_100GeV}
     \end{subfigure}
     \hfill
     \begin{subfigure}[b]{0.45\textwidth}
         \centering
         \includegraphics[width=\textwidth]{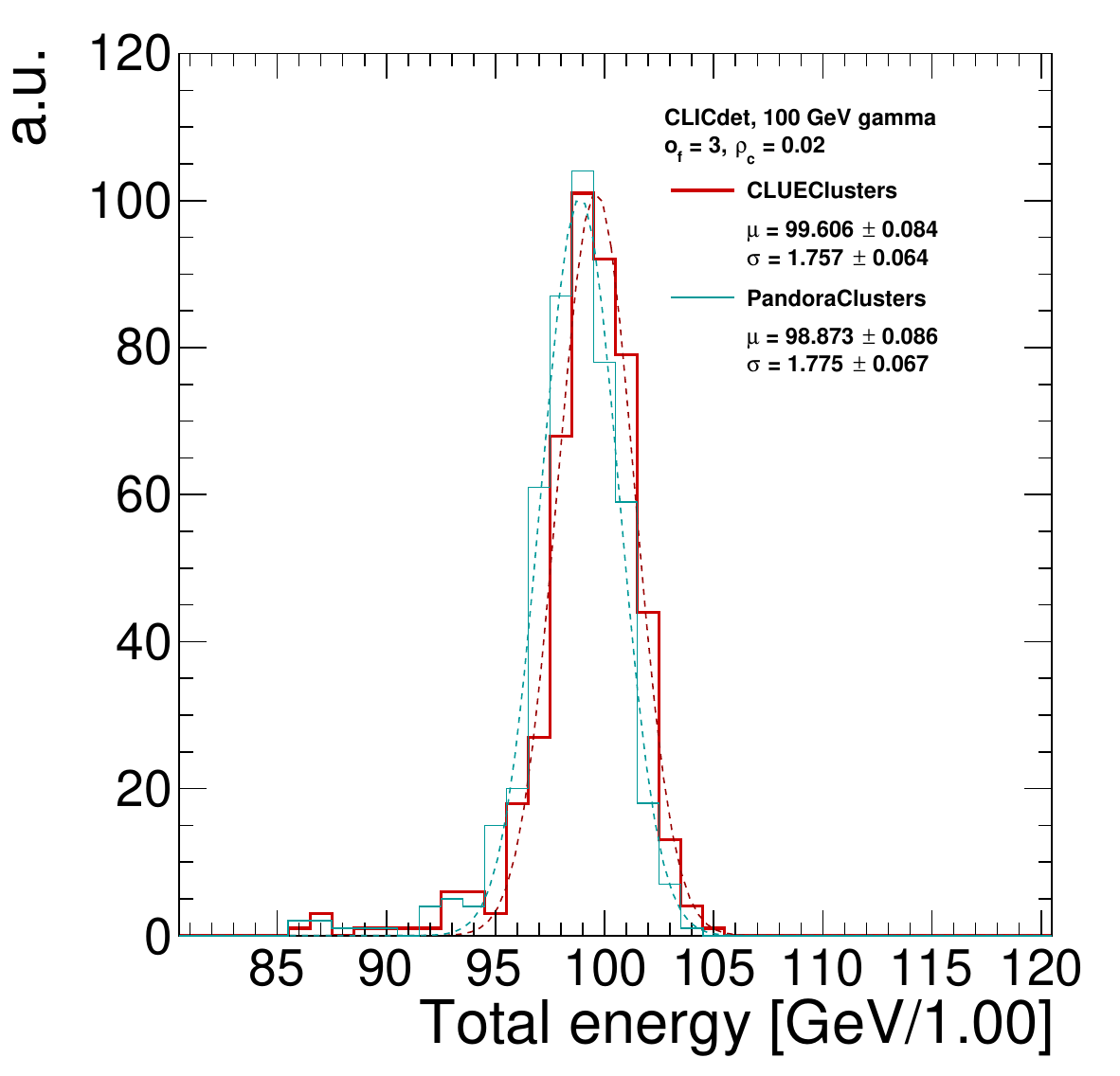}
         \label{fig:clicdet_energy_100GeV}
     \end{subfigure}
     \caption{
     In the left plots, the average number of CLUE's followers (full rhombus), seeds (full circles), and outliers (empty squares) are shown, while in the right plots, the comparison between the clustered energy using the CLUE algorithm (thicker red line) and the \texttt{PandoraPFA} toolkit is presented. Results are produced in the upper (lower) row with single gamma events generated from the inner surface of the CLICdet calorimeter with monochromatic energy of 10 (100)~\GeV. The CLUE input parameters are fixed at $o_f = 3$, and $\rho_c = 0.02$. }
     \label{fig:clicdet_performance}
\end{figure}

The left panel of Figure~\ref{fig:clicdet_performance} displays the average number of CLUE's followers, seeds, and outliers as a function of the layer number, while the right panel shows the distribution of the total energy collected in the clusters. The top row corresponds to 10\GeV single gamma events, while the bottom row shows the results for 100\GeV single gamma events. As expected, due to the higher energy of the 100\GeV showers, the peak of the number of followers shifts to higher layers, consistent with the expected electromagnetic shower profile.

The clustered energy distribution is compared to that obtained from the \texttt{PandoraPFA} C++ Software Development Kit~\cite{PandoraPFA}, which uses multiple algorithms to reconstruct and to identify different types of particles in an event. The \texttt{PandoraPFA} package also has a dedicated calibration procedure that uses software compensation. Thus, it is acknowledged by the authors that this is not a completely equitable comparison, but the results show that the \texttt{k4CLUE} package is capable of reconstructing particle showers at higher energies without requiring any changes to the input parameters and that the energy linearity and resolution achieved by \texttt{k4CLUE} are comparable to those obtained by the official particle flow algorithm.

%% file: noble_liquid_calo.tex
\subsection{Performance for Noble Liquid Calorimeter}\label{sec:lar}

The detector simulation and local reconstruction studies presented for the Noble Liquid Calorimeter were conducted using the \texttt{k4RecCalorimeter} package developed within the FCC Software and that specifically contains \texttt{Key4hep} framework components for calorimeter reconstruction~\cite{k4RecCalorimeter_github}.

\begin{figure}
     \centering
     \begin{subfigure}[b]{0.45\textwidth}
         \centering
         \includegraphics[width=\textwidth]{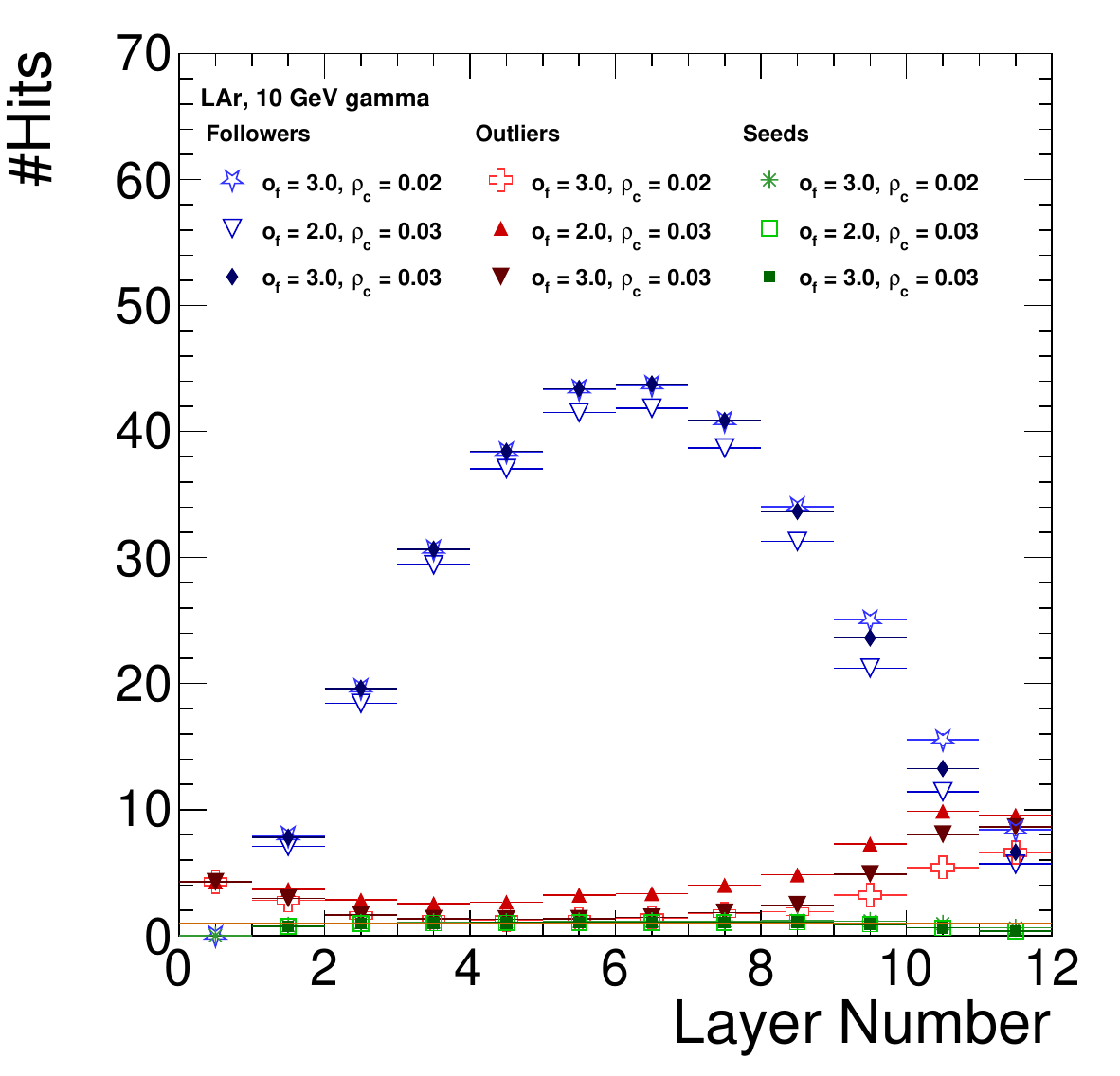}
         \label{fig:lar_sub_tuning}
     \end{subfigure}
     \hfill
     \begin{subfigure}[b]{0.45\textwidth}
         \centering
         \includegraphics[width=\textwidth]{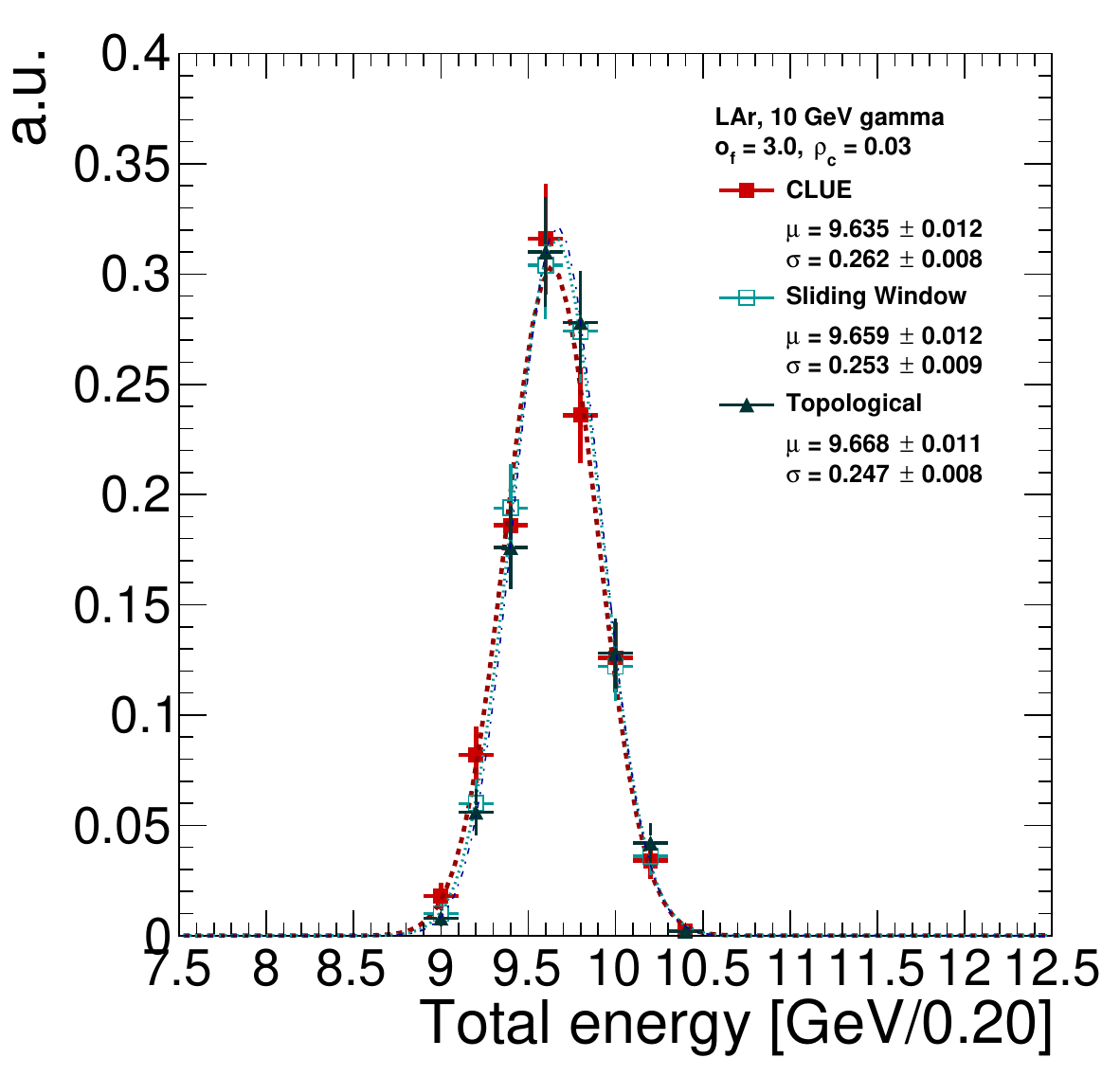}
         \label{fig:lar_performances_no_noise}
     \end{subfigure}
     \caption{
     The left plot shows the average number of followers (blue), outliers (red) and seeds (green) for different combinations of $\rho_c$ and $o_f$ using the CLUE algorithm. The right plot compares the total clustered energy obtained using CLUE (full red squares), sliding window algorithm (empty cyan squares), and topological clustering (full blue triangles), along with the Gaussian fit to the distribution. These results are obtained using single gamma events with monochromatic energy of 10\GeV generated from the interaction point. The input parameters for CLUE in the left plot are fixed at $o_f = 3$ and $\rho_c = 0.03$.
     }
     \label{fig:lar_tuning}
\end{figure}

Similarly to the previous calorimeters, the critical distance, $d_c$, in the Noble Liquid Calorimeter is chosen to take into account the granularity of the detector's geometry and to ensure that neighboring cells are contained is taken into account. Therefore, this is set to approximately two times the size of a single cell, $d_c = 40.0\mm$. Various combinations of $\rho_c$ and $o_f$ were also tested for this calorimeter, and the average number of followers, outliers and seeds is displayed in Figure~\ref{fig:lar_tuning}~(left) for 500 single gamma events generated on the interaction point within the entire detector with monochromatic energy of 10\GeV. Similar conclusions to those discussed for the previous calorimeters can be applied to the Noble Liquid Calorimeter.

FCCSW already includes two types of algorithms to cluster energy deposits in the Noble Liquid Calorimeter: the sliding window algorithm that produces clusters of a fixed size in $\Delta\eta\times\Delta\phi$ and a constant size in radius $r$, and the topological clustering that groups adjacent cells according to their energy deposits starting from a seed cell~\cite{LAr_FCChh_2019}.
Figure~\ref{fig:lar_tuning}~(right) compares the distribution of the clustered energy and its Gaussian fit for the two FCCSW algorithms and the CLUE algorithm. Despite the differences in the clusterization procedure, the energy peak and resolution performances are similar. 

\begin{figure}
     \centering
     \begin{subfigure}[b]{0.45\textwidth}
         \centering
         \includegraphics[width=\textwidth]{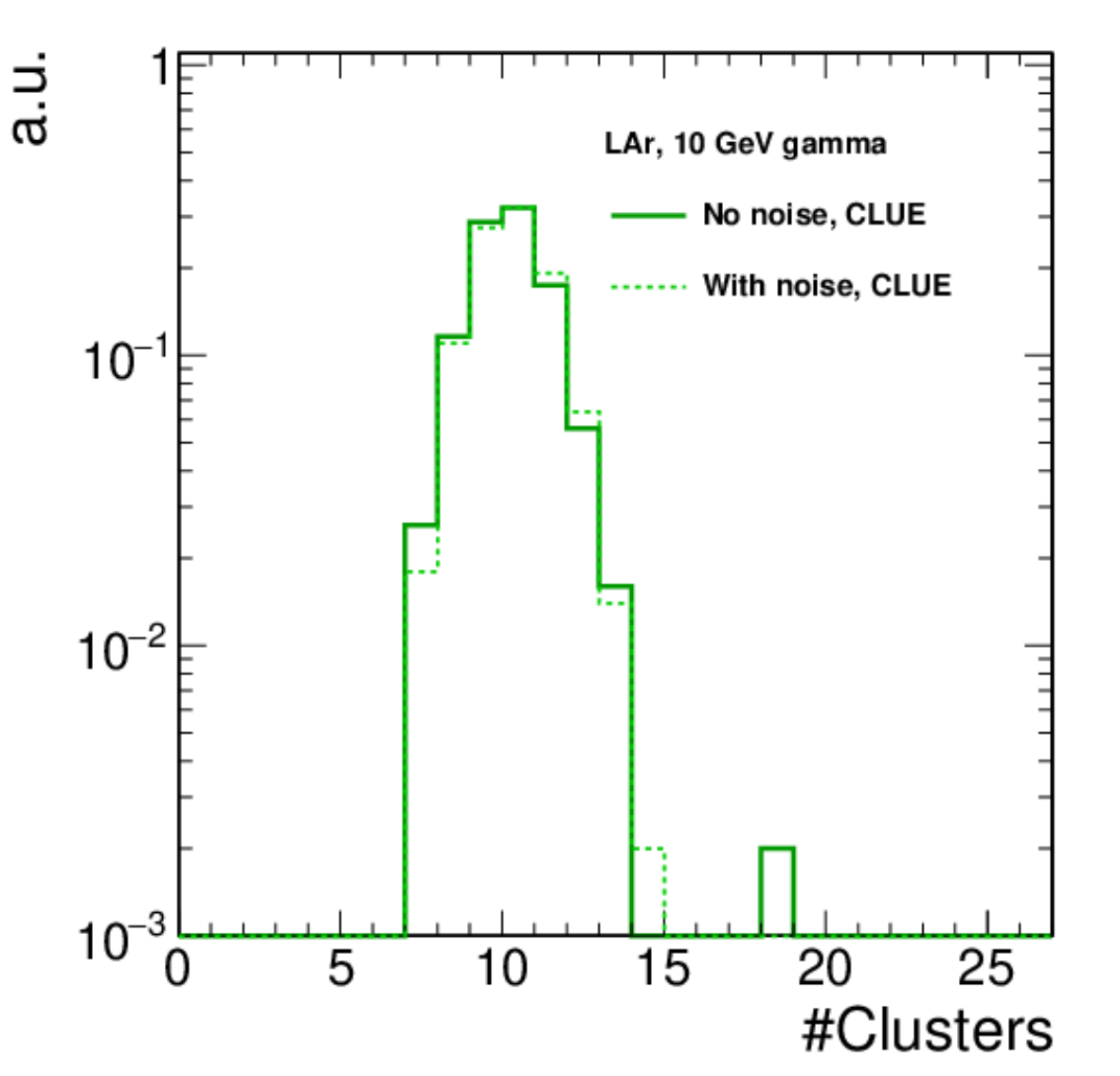}
     \end{subfigure}
     \hfill
     \begin{subfigure}[b]{0.45\textwidth}
         \centering
         \includegraphics[width=\textwidth]{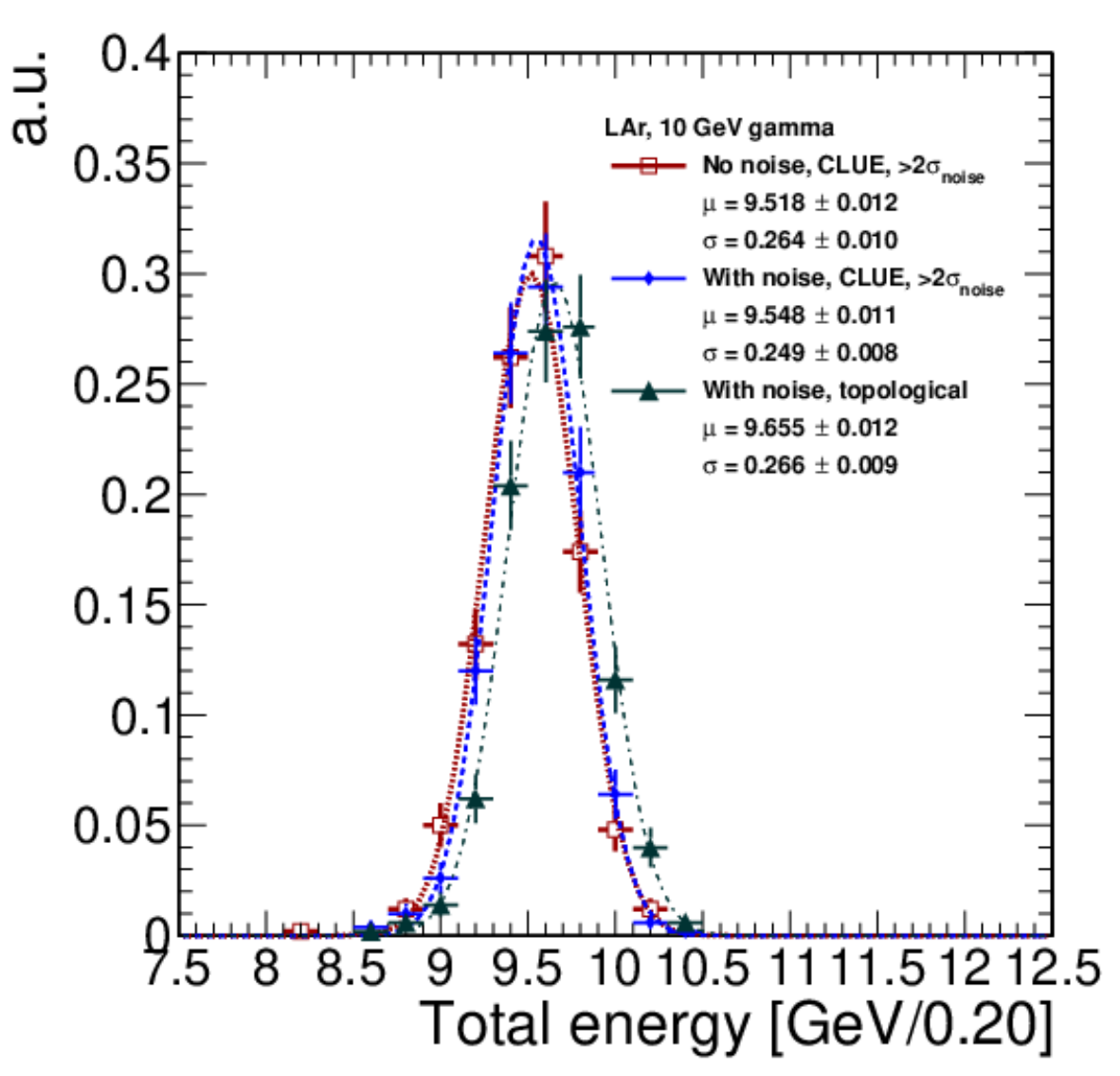}
     \end{subfigure}
     \caption{On the left, the distribution of the number of clusters reconstructed by CLUE is presented without (solid line) and with (dashed line) the noise simulation. On the right, the clustered energy distribution is shown for CLUE without (empty red squared) and with (full blue rhombus) the noise simulation and for the topological clustering algorithm (full green triangles).
     }
     \label{fig:lar_performance}
\end{figure}

To evaluate the clustering performance of the CLUE algorithm under noisy conditions, we simulated cell noise level between 0.5 to 2.0~\MeV in the detector depending on the longitudinal layer as described in~\cite{LAr_FCCee_2022}. We then applied a pre-filter on the calorimeter cell energy at $2\sigma_{noise}$, similar to the one applied in the clusterization process of the topological algorithm. The results are shown in Figure~\ref{fig:lar_performance}, where the distribution of the number of clusters reconstructed by CLUE is unaffected by the noise, and the clustered energy is compared before and after the noise introduction and to the topological clustering algorithm. This demonstrates the robustness of the CLUE algorithm under noisy conditions.

\begin{figure}
     \centering
     \begin{subfigure}[b]{0.65\textwidth}
         \centering
         \includegraphics[width=\textwidth]{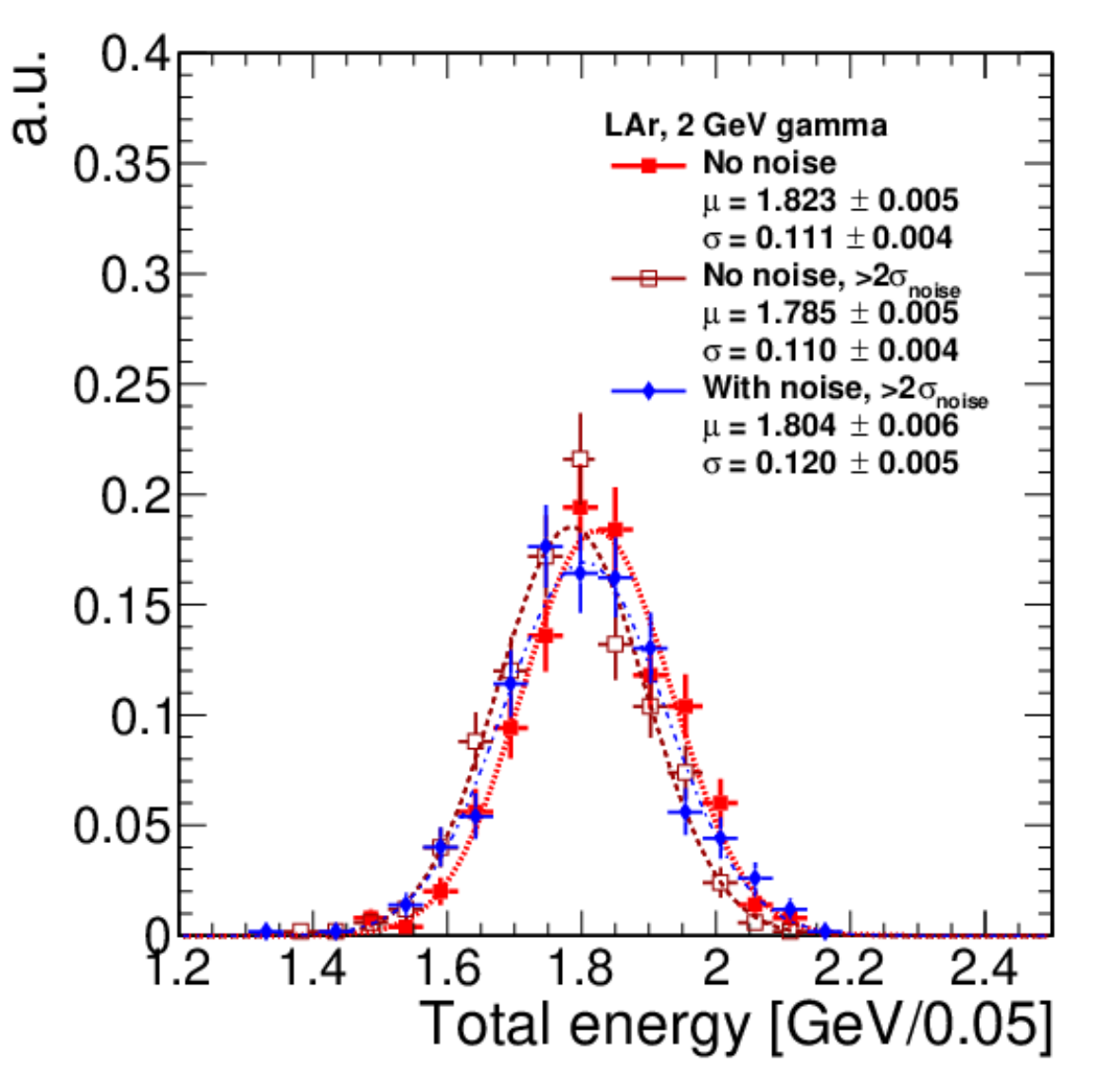}
         \label{fig:lar_seeds_tuning}
     \end{subfigure}
     \caption{
     The clustered energy distribution is shown for CLUE without (full red squares) and with (empty dark red squares) a pre-filter applied and including the noise (full blue rhombus). These results are obtained using single gamma events with monochromatic energy of 2\GeV generated from the interaction point. The input parameters for CLUE are fixed at $o_f = 3$ and $\rho_c = 0.03$.
     }
     \label{fig:lar_lower_en}
\end{figure}

As a final assessment, we evaluated the clustering performance of CLUE for low-energy gamma particles, which are important for certain flavor physics studies at the Z peak. Figure~\ref{fig:lar_lower_en} displays the comparison of clustered energy by CLUE, with and without a pre-filter on the calorimeter cell energy at $2\sigma_{noise}$ and including the noise simulation. The results confirm that the algorithm is robust in this scenario too.

%% file: computing_time.tex
\subsection{Execution time}\label{sec:timing}

We tested the computational performance of CLUE, comparing it to the topological clustering and sliding window algorithms described in Sec.~\ref{sec:lar}. For this evaluation, 500 events of single gamma were generated at the interaction point within the entire detector, each with a monochromatic energy of 10\GeV. The input parameters for CLUE remained the same as described.

\begin{figure}
     \centering
     \begin{subfigure}[b]{0.45\textwidth}
         \centering
         \includegraphics[width=\textwidth]{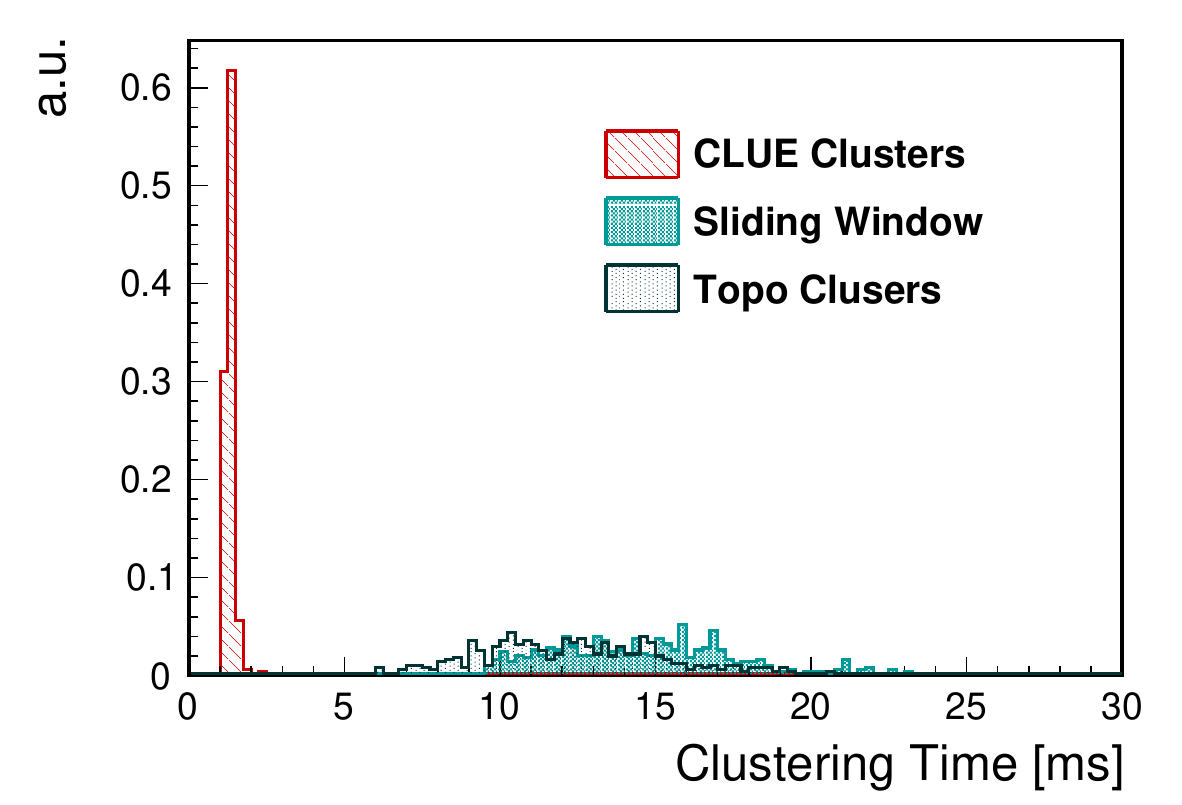}
     \end{subfigure}
     \hfill
     \begin{subfigure}[b]{0.45\textwidth}
         \centering
         \includegraphics[width=\textwidth]{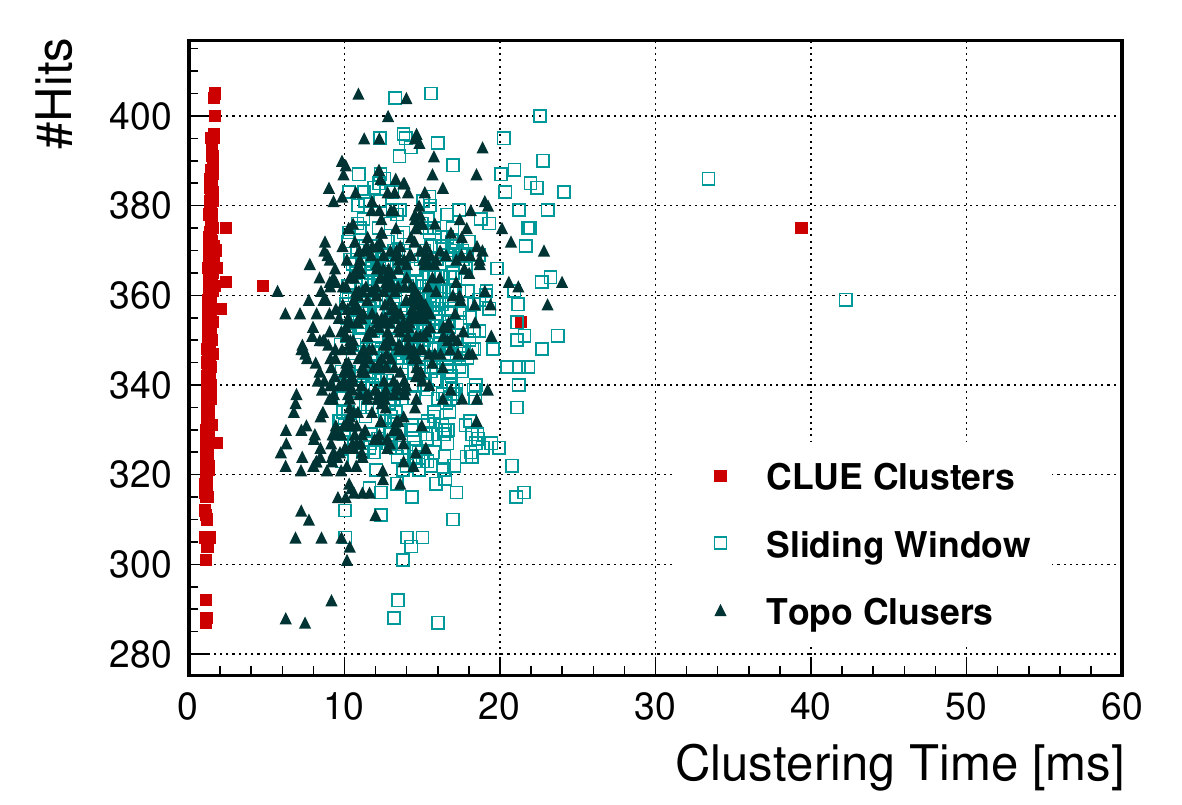}
     \end{subfigure}
     \caption{The total clustering time per event (left) and the number of hits as a function of the clustering time (right) is shown using CLUE (red full squares), sliding window algorithm (cyan empty squares), and topological clustering (blue triangle).
     }
     \label{fig:timing_performance_lar}
\end{figure}

In Fig.~\ref{fig:timing_performance_lar}, we present the clustering time per event for the three algorithms on the left, and, on the right, we display the number of hits as a function of the clustering time. CLUE demonstrates impressive timing capabilities, outperforming the other algorithms by completing the task in about a tenth of the time, regardless of the number of input hits.

It is worth noting that the timing performance of CLUE is calculated using only the CPU version of the code. As the \texttt{Key4hep} framework evolves to incorporate GPU utilization, CLUE's implementation on GPU could further enhance its performance.

